\documentclass[12pt]{article}
\usepackage[margin=1cm]{geometry}
\usepackage{fullpage}
\usepackage{setspace,verbatim}
\usepackage{amsmath,amsfonts,amssymb,amsthm,cite,graphicx,subfigure,color}
\numberwithin{equation}{section}

\newcommand{\calO}{{\cal O}}

\begin{document}
\reversemarginpar
\thispagestyle{empty}

\begin{center}

\vspace{10cm}
{\LARGE{\bf Scalar Boundary Conditions in Lifshitz Spacetimes}}

\vspace{1cm}

Cynthia Keeler

\vspace{0.8cm}

{\it Department of Physics, University of Michigan,\\
Ann Arbor, MI-48109, USA \\}

\vspace{0.6cm}

{\tt keelerc@umich.edu}\\

\vspace{2cm}
\end{center}
\begin{abstract}
We investigate the conditions imposable on a scalar field at the boundary of the so-called Lifshitz spacetime which has been proposed as the dual to Lifshitz field theories.  For effective mass squared between $-(d+z-1)^2/4$ and $z^2-(d+z-1)^2/4$, we find a one-parameter choice of boundary condition type.  The bottom end of this range corresponds to a Breitenlohner-Freedman bound; below it, the Klein-Gordon operator need not be positive, so we cannot make sense of the dynamics.  Above the top end of the range, only one boundary condition type is available; here we expect compact initial data will remain compact in the future.
\end{abstract}

\pagebreak
\setcounter{page}{1}
\section{Introduction}
Recent interest in extending gauge-gravity duality to theories with non-relativistic scaling has led to questions about the nature of holography in spacetimes with formally degenerate boundaries.  Lifshitz- and Schroedinger-dual spacetimes both have this feature, as do the warped AdS spacetime solutions in topologically massive gravity.  In fact, any spacetime which is dual to a non-relativistic field theory must have a degenerate boundary; the non-relativistic symmetry means that the time and space components of the metric do not scale the same with respect to the radial coordinate.  Thus, there will be no single overall conformal scaling which we could use to define a nondegenerate boundary metric.

There are several known ways of dealing with this issue.  The most common way is to simply define a metric for the field theory at a cutoff surface near the edge of the bulk.  This method is calculationally practical but does not provide for a theoretical understanding of the nature of the degenerate boundary itself.  However, as we will show, it is still possible to gain further understanding of the available boundary conditions via a cutoff approach.

In this paper we will concentrate on the Lifshitz spacetime as defined in \cite{Kachru2008}. We will not address the Schroedinger spacetime; for references which do, see \cite{Guica2010,Guica2011a,VanRees2011}.  Ho\v{r}ava-Lifshitz gravity (distinct from the Lifshitz spacetime first proposed in \cite{Kachru2008}) also does not have a well-defined conformal boundary; motivated by the preferred time slicing present in Ho\v{r}ava-Lifshitz gravity, \cite{Horava2009} proposed that we define the boundary for all spacetimes by peeling off a different factor for each piece of the metric, chosen to ensure that the pieces remain finite as the radius approaches infinity.

In a similar vein, \cite{Ross2009,Ross2011} define a stress-tensor complex at the boundary of Lifshitz spacetime by applying an appropriate conformal factor to each portion of the tensor. Additionally, \cite{Ross2011} provides a definition of an asymptotically locally Lifshitz spacetime.  Unfortunately, in order to preserve the asymptotic conditions, it becomes necessary to turn off some possible modes when doing holography.  \cite{Baggio2011} sets boundary conditions in a different manner, by first requiring that all divergences be cancellable by local counterterms.  \cite{Mann2011} builds a stress-energy tensor at the boundary specifically for spacetimes with $z=2$, again imposing a set of boundary conditions a priori.  \cite{Copsey} studies perturbations in a particular Hamiltonian formulation, again imposing boundary conditions and thus limiting the available solutions. \cite{Hoyos2010} considers the effective of the null energy condition on causality at Lifshitz boundaries.

In this paper, we will explore the possible boundary conditions for a scalar field living on a background Lifshitz spacetime, following the approach of \cite{Ishibashi2003,Ishibashi2004}.  In \cite{Ishibashi2003}, the authors provide a unique prescription for studying the possible boundary conditions for perturbations on a stably causal, static spacetime which are compatible with ``good dynamics''.  By ``good dynamics'', they mean it is possible to extended initial data for a wave equation on a static time slice to a solution throughout the spacetime, while also maintaining
\begin{itemize}
\item local agreement with wave equation
\item a positive conserved energy
\item time translation and reflection.
\end{itemize}
When they additionally impose a particular convergence condition first defined in \cite{Wald1980}, \cite{Ishibashi2003} shows that their prescription for finding consistent boundary conditions is unique.

In \cite{Ishibashi2004}, the same authors explore this boundary condition prescription for global anti-de Sitter space.  They find that for a low-mass range just above the Breitenlohner-Freedman bound, there is a full one-parameter range of allowed boundary conditions, ranging from Dirichlet to Robin to Neumann conditions. These mixed boundary conditions were studied in \cite{Amsel2006} and subsequent papers, and associated with multi-trace perturbations of the dual CFT.

In this paper we will apply the approach of \cite{Ishibashi2003} to the Lifshitz spacetime with flat slicing,
\begin{equation}
ds^2=-\frac{dt^2}{r^{2z}}+\frac{d\vec{x}^2}{r^2}+\frac{dr^2}{r^2},
\end{equation}
which is both static and stably causal.  We will not propose any specific action for which this spacetime is the solution; rather we only consider scalar perturbations living on this background.

We privilege the static time slicing, constructing a spatial boundary for each time slice; alternatively, we could imagine the boundary of the spacetime as being at a cutoff surface inside spatial infinity.  Either approach allows us to consider boundary conditions for a scalar field at spatial infinity on each time slice. In the AdS case where $z=1$, we reproduce the alternate set of boundary conditions as found in \cite{Ishibashi2004}. When $z>1$, we again find a one-parameter family of conditions available in the low-mass range just above a Breitenlohner-Freedman-type bound.  Above this range, only one type of boundary condition is consistent.

Concurrently with this paper, \cite{Andrade2012} studied the same problem and found a similar constraint, but also discovered a novel instability at the upper end of the low-mass range indicated here.  Later \cite{Andrade2013} also explored the boundary conditions imposable on metric fluctuations.

In Section \ref{WIreview}, we review the approach to studying boundary conditions presented in \cite{Ishibashi2003}.  In Section \ref{AdSpoincare}, we reproduce the global analysis of the scalar field in AdS done in \cite{Ishibashi2004} in the Poincar\'e patch.  In Section \ref{Lifsection}, we study the boundary conditions for scalar fields in Lifshitz spacetimes for rational scaling parameters $z>1$. In Section \ref{conclusions}, we discuss the comparison of our results to previous work, and consider future extensions.

\section{Review of Wald and Ishibashi Procedure}\label{WIreview}
We will now review the procedure described in \cite{Ishibashi2003}.  We are interested in studying the behavior of fields $\phi$ solving particular wave equations on a spacetime. If a spacetime is globally hyperbolic, then we do not need to impose boundary conditions at the edge of spacetime for these fields; just knowing the initial data $\phi_0,$ $\dot{\phi}_0$ on one Cauchy surface is sufficient to evolve the field throughout the spacetime.

For spacetimes that are not globally hyperbolic, there is no such Cauchy surface in the spacetime.  If we want to evolve initial data throughout the spacetime, we must include additional information about what can come in through the boundary at spatial infinity or at singularities. If the spacetime is stably causal, we can impose boundary conditions that produce ``good dynamics''.

We impose stable causality because spacetimes which are not stably causal are arbitrarily close to having closed timelike curves.  Thus, the perturbations due to backreaction could produce causal violations, and prevent us from evolving our initial data in a sensible manner.

Stable causality is equivalent to the existence of a well defined time function which increases along all timelike curves in the spacetime.  Thus, all stably causal spacetimes have well defined time-slicings.  If a spacetime is additionally static, then each time-slice will be equivalent to any other; we are free to consider the physics on one particular slice $\Sigma$.  By studying the boundary conditions for fields $\phi$ on a given timeslice $\Sigma$, we thus deduce possible boundary data for $\phi$ on the spacetime as a whole.

The procedure developed in \cite{Ishibashi2003} finds the consistent boundary data imposable for a field $\phi$ solving a wave equation on a static stably causal spacetime, such that these boundary data allow ``good dynamics'', as defined in the introduction, for the field $\phi$ in the full spacetime.
It involves essentially five steps:
\begin{itemize}
\item  Pick a static direction $t$ which gives a timelike slicing with slice $\Sigma$.
\item Find the operator $A$ which describes the spatial portion of the wave equation solved by $\phi$ and describe its Hilbert space.
\item Delineate the positive self-adjoint extensions $A_U$ of $A$.
\item Use $A_U$ to define $\phi_t$ which solves the full wave equation throughout the entire spacetime.
\item Analyze the boundary conditions obeyed by $\phi_t$ for a given extension $A_U$.
\end{itemize}
These steps provide a precise means for extending the operator $A$ to an action $A_U$ on the spatial boundary of $\Sigma$, which is well-defined since it is a single time-slice.  Additionally, they allow us to examine the domain of the new $A_U$, and thus the boundary conditions which the $\phi_t$ will obey as we propogate the data $\phi_0$ into future time slices.

Note that positivity of $A_U$ refers to the condition $\langle \phi |A_U \phi \rangle >0$ for all $\phi$ in the Hilbert space.  Negative norm would indicate an instability in the future evolution of $\phi_t$, so we wish to avoid it.
We now proceed to explore each of these steps in more detail.

\subsection{Finding the operator $A$}

First we pick out a static direction $t$.  Some spacetimes may allow multiple choices of static direction; in the case of AdS, we may choose either global time or Poincar\'e time.  We expect that boundary condition results should not depend on the choice of slicing.  In any case, picking a static time direction $t$ privileges a particular time slicing with equivalent slices $\Sigma$. We can write the wave equation solved by the scalar field as
\begin{equation}\label{Aeqn}
\frac{\partial^2}{\partial t^2}\phi=-A \phi,
\end{equation}
for some operator $A$ that acts only on $\Sigma$.  The domain of $A$ is the set of smooth functions of compact support which live on
$\Sigma$.  We treat $A$ as an operator acting on the Hilbert space of square integrable functions on $\Sigma$, but with volume measure
$V^{-1}d\Sigma$, where
\begin{equation}
V\equiv (-t^\mu t_\mu)^{1/2}
\end{equation}
encodes the effect of the time components of the metric.

\subsection{Finding positive self-adjoint extensions of $A$}
Given the operator $A$, we will need to enumerate its positive self-adjoint extensions. We will additionally need to insist that $A$ itself to be positive and symmetric; it is this restriction which will give us the Breitenlohner-Freedman bound on the mass of a scalar.  Here we will only summarize the procedure for finding these self-adjoint extensions; in \cite{Ishibashi2003}, the authors explain why such extensions will be related to the available boundary conditions which preserve ``good dynamics''.

First, we find the size of the solution spaces $\mathcal{K}_\pm$ whose elements solve
\begin{equation}\label{Aieqn}
A\psi_\pm=\pm i\psi_\pm,
\end{equation}
and are also square integrable under the measure $V^{-1}d\Sigma$.

If $\mathcal{K}_\pm$ are both empty, then there is a unique extension, given by $\bar{A}$, the closure of $A$ (this is the Friedrichs extension).  Thus there is a unique choice of boundary condition in this case.

Conversely, if $\mathcal{K}_\pm$ are the same dimension, then we find the maps $U: \mathcal{K}_+ \to \mathcal{K}_-$ such that $||\phi_+||=||U\phi_+||$.  Given a choice of $U$, we then define a particular self adjoint extension via
\begin{equation}
A_U \phi = \bar{A} \phi_0 +i \phi_+-i U\phi_+.
\end{equation}
Note that the domain of $A_U$ is larger than that of $\bar{A}$:
\begin{equation}\label{AUdomain}
\text{Dom} (A_U)=\left\{\phi_0+\phi_+ + U\phi_+ | \phi_0 \in \text{Dom}(\bar{A}), \phi_+ \in \mathcal{K}_+ \right\}.
\end{equation}
Lastly, we check positivity of proposed $A_U$.  We have one positive self adjoint extension for every $U$ which produces a positive $A_U$.

\subsection{Finding $\phi_t$}
We want to find time-dependent solutions $\phi_t$ to the equation (\ref{Aeqn}). We can define
\begin{equation}\label{phitequation}
\phi_t=\cos(A_U^{1/2} t)\phi_0 + A_U^{-1/2}\sin(A_U^{1/2} t)\dot{\phi}_0,
\end{equation}
where $\phi_0$ and $\dot{\phi}_0$ represent the initial data on our reference time slice $\Sigma_0$.  As shown in \cite{Ishibashi2003},
if $\phi_0$ and $\dot{\phi}_0$ are smooth functions of compact support on $\Sigma_0$, then $\phi_t$ will be a smooth solution to (\ref{Aeqn}) throughout the spacetime.  $\phi_t$ matches the initial data on our reference time slice, and produces ``good dynamics''.

Lastly, we study the boundary conditions satisfied by $\phi_t$ defined as in (\ref{phitequation}).  Thus we relate the choice of map $U$ to a choice of boundary condition on the field $\phi$.

\section{Scalars in AdS}\label{AdSpoincare}

We now apply the procedure from \cite{Ishibashi2003} as reviewed in Section \ref{WIreview} to the case of a scalar field in AdS spacetime on the Poincar\'e patch.  We will reproduce results in \cite{Ishibashi2004}, which studied fields on AdS spacetime in global coordinates.  \cite{Ishibashi2004} did not restrict its results to scalar fields; rather, the authors reduced equations for graviton modes and gauge modes to effective scalar equations by exploiting the $SO(2,d-2)$ symmetry.  Here, we will concentrate only on scalar fields solving the Klein-Gordon equation; we leave analysis of other modes to future work.

In particular we use the Poincar\'e patch because the Lifshitz spacetime we wish to study does not have a good global coordinate system; instead only the analogue of the Poincar\'e patch is available.  We should also note that AdS global time is not equivalent to AdS Poincar\'e time; therefore some of the details of our analysis will differ, but the end result describing the boundary conditions at the spatial boundary will be the same.

Specifically, we work with $AdS_{d+1}$ written as
\begin{equation}\label{AdSmetric}
ds^2_{d+1}=-\frac{dt^2}{r^2}+\frac{d\vec{x}^2}{r^2}+\frac{dr^2}{r^2}.
\end{equation}
We study the Klein-Gordon equation for a massive scalar in this background, which becomes
\begin{equation}\label{AdSKG}
-r^2\partial_t^2\phi+r^{d+1}\partial_r\left[r^{1-d}\partial_r \phi\right]+r^2 \vec{\nabla}\cdot\vec{\nabla}\phi-m_0^2\phi=0.
\end{equation}
The Hilbert space consists of functions on spatial slices $\Sigma$ which are normalizable under the norm
\begin{equation}\label{AdSnorm}
\langle \phi_2 | \phi_1 \rangle = \int_{\Sigma}\phi_2^* \, \phi_1 \, r^{1-d} \, dr \, d\vec{x}_{d-1}.
\end{equation}
Since the wave equation is fully separable, we expand $\phi$ as an integral over plane waves labelled by $\vec{k}$; additionally, we will multiply by a power of $r$ designed to remove the single $r$-derivative term in (\ref{AdSKG}).  Thus we define $\psi_k(t,r)$ via
\begin{equation}\label{AdSpsidef}
\phi= r^{\frac{d-2}{2}}\int dk \, c_{\vec{k}}\, e^{i\vec{k}\cdot\vec{x}} \, \psi_k(t,r),
\end{equation}
where $c_{\vec{k}}$ depends only on $\vec{k}$ and tells us the shape of the wave packet in the $\vec{x}$ directions.  Since we are only interested in behavior at the $r=0$ boundary, we will only need to consider boundary conditions on the $\psi_k$.  We can now find the operator $A$ by rewriting the Klein-Gordon equation (\ref{AdSKG}) as
\begin{align}
\partial_t^2 \psi_k &=-A\,  \psi_k,\notag\\
\label{AdSAdef}
A &\equiv -\partial_r^2+\frac{\nu^2-\frac{1}{4}}{r^2}+k^2,
\end{align}
where $\nu^2 \equiv m_0^2+(d/2)^2$.  The Breitenlohner-Freedman bound is saturated for $\nu=0$; $A$ is also only positive for masses above this bound.  Accordingly we will only study $\nu\geq 0$.

Under the redefinition (\ref{AdSpsidef}), the $r$ portion of the norm (\ref{AdSnorm}) becomes simply
\begin{equation}\label{AdSnorm2}
\langle \psi_2|\psi_1\rangle_r =\int dr \, \psi_{2,k}^* \, \psi_{1,k}.
\end{equation}

We will now follow the procedure outlined in Section \ref{WIreview}.  First we will study the eigenvalue equation (\ref{Aieqn}) with the operator $A$ as in (\ref{AdSAdef}) under the Hilbert space of functions normalizable under (\ref{AdSnorm2}).  We then use these solutions to construct the self adjoint extensions $A_U$ of $A$, and study which boundary conditions these extensions correspond to.

\subsection{Solving $A\psi=\pm i \psi$}
We wish to study solutions of
\begin{equation}
A\psi=\lambda\psi,
\end{equation}
with $A$ as in (\ref{AdSAdef}).  We can rewrite this equation as
\begin{equation}\label{AdSeigen}
-r^2\partial_r^2\psi+\left(\nu^2-\frac{1}{4}\right)\psi+\left(k^2-\lambda\right) r^2 \psi=0.
\end{equation}
The solutions to this equation can be written in terms of Hankel functions, which are given in terms of the usual Bessel functions as
\begin{align}
H_{1,\nu} (a r)&= J_{\nu}(a r) + i Y_{\nu} (a r).\notag \\
H_{2,\nu} (a r)&= J_{\nu} (a r) - i Y_{\nu} (a r).
\end{align}
Their large $r$ behavior is
\begin{align}
H_{1,\nu}(a r) &= \left(\frac{2}{\pi a r}
\right)^{1/2}
e^{i \left[a r -\frac{\nu \pi}{2}-\frac{\pi}{4}
\right]} ,
\notag \\\label{largerHankel}
H_{2,\nu}(a r) &= \left(\frac{2}{\pi a r}
\right)^{1/2}e^{-i \left[a r -\frac{\nu \pi}{2}-\frac{\pi}{4}
\right]},
\end{align}
for $|\arg a|<\pi$.
The solutions to (\ref{AdSeigen}) are
\begin{equation}\label{psilambda}
\psi_\lambda (r)= C_1 \sqrt{r}H_{1,\nu}(r\sqrt{\lambda-k^2})+C_2 \sqrt{r}H_{2,\nu}(r\sqrt{\lambda-k^2}),
\end{equation}
where by $\sqrt{\lambda-k^2}$ we mean the root with positive imaginary component. This is just a parametrization choice and is always available for $\lambda=\pm i$, the case we are interested in. We make this choice so $H_1$ will be exponentially damped near $r\to \infty$, while $H_2$ will blow up there.

For $0\leq \nu < 1$, both $\sqrt{r}Y_{\nu}$ and $\sqrt{r}J_{\nu}$ are normalizable near $r=0$ under the norm (\ref{AdSnorm2}).  Consequently, as we can see from the large $r$ behavior, the solution $H_1$ will be square integrable under the norm (\ref{AdSnorm2}).  The solution $H_2$, however, blows up exponentially near $r\to \infty$ and thus is not square integrable.

 For $\nu\geq 1$, neither solution is normalizable, because only $\sqrt{r}J_{\nu}$ is normalizable near $r=0$, so both solutions $H_1$ and $H_2$ are disallowed.  We cannot consider $\sqrt{r}J_{\nu}$ itself either, because it will be a combination of $H_1$ and $H_2$, and will thus blow up near $r \to \infty$.

Thus, for $0\leq \nu <1$, we have one (linearly independent) solution to (\ref{Aieqn}) for every value of $k$ and each of $\lambda=\pm i$, and thus the dimension of both $\cal{K}_\pm$ is $1$.
For $\nu\geq 1$, $\cal{K}_\pm$ are both empty.  These results match with those of \cite{Ishibashi2004}.

\subsection{Finding the extensions $A_U$}
We now use the results of the previous section to build the self-adjoint extensions $A_U$.  The arguments follow similarly to those in \cite{Ishibashi2004}.

First, for the case $\nu\geq 1$, the only available extension of $A$ is just the closure of $A$, $\bar{A}$.  That is, for $\nu \geq 1$, we simply extend the action of $A$ on each slice $\Sigma$ to the spatial boundary of $\Sigma$ at $r=0$.

For perturbations whose effective $\nu$ satisfies $0\leq \nu <1$, both $\cal{K}_+$ and $\cal{K}_-$ are one dimensional.  Specifically they are spanned by $\psi_\pm$, defined from (\ref{psilambda}) as
\begin{equation}
\psi_\pm=\psi_{\lambda=\pm i}= C_\pm \sqrt{r}H_{1,\nu}\left(r\sqrt{\pm i-k^2}\right),
\end{equation}
where $\sqrt{\pm i-k^2}$ refers to the square root with positive imaginary component, and the $C_\pm=C$ are positive real constants chosen such that $||\psi_\pm||=1$ under the norm (\ref{AdSnorm2}).  It can be shown that $C$ is the same for both $\psi_+$ and $\psi_-$.

As the sets $\cal{K}_\pm$ are one dimensional, we can write all maps which take $\cal{K}_+ \rightarrow \cal{K}_-$ in the form
\begin{equation}\label{phasemap}
U \psi_+ =e^{i\alpha} \psi_-,
\end{equation}%
for $-\pi<\alpha\leq \pi$.  In other words, we can only introduce a phase into the map, but the phase can take any value.  Note that the transformation $U$ takes the element $\psi_+$ of $\cal{K}_+$ and associates it with a particular element, $e^{i\alpha}\psi_-$, in the space $\cal{K}_-$.  Our only definition of the action of $U$ is as given in (\ref{phasemap}).

For a particular choice $\alpha$, the map corresponds to the extension of $A$ whose operation on $\psi$ is given by
\begin{equation}\label{Aextension}
A_\alpha \psi= \bar{A} \psi_0+ i \psi_+ - i e^{i\alpha} \psi_-.
\end{equation}

\subsection{Interpreting the boundary conditions}

Again following \cite{Ishibashi2004}, we will need to understand the asymptotic behavior of
\begin{equation}
\psi_\alpha \equiv \psi_+ + e^{i\alpha} \psi_- =C \sqrt{r}H_{1,\nu}\left(r \sqrt{i-k^2}\right)+e^{i\alpha}C\sqrt{r}H_{1,\nu}\left(r\sqrt{-i-k^2}\right)
\end{equation}
near the boundary $r=0$, for $0\leq \nu< 1$.  Expanding $\psi_\alpha$ near $r=0$ for $\nu$ not a half integer, we find
\begin{align}\label{psialphanear0}
\psi_\alpha C^{-1}=&\quad r^{1/2-\nu}
\left(
\frac{-i\Gamma(\nu)2^\nu}{\pi\sqrt{i-k^2}^\nu}
\right)
\left[
1+e^{i\alpha}\left(\frac{\sqrt{i-k^2}}{\sqrt{-i-k^2}}\right)^\nu
\right]
\\
&+r^{1/2+\nu}\left(
\frac{\sqrt{i-k^2}^\nu}{2^\nu}
\right)
\left(
\frac{1}{\Gamma(1+\nu)}
-\frac{i \cos (\pi\nu)\Gamma(-\nu)}{\pi}
\right)
\left[
1+e^{i\alpha}\left(\frac{\sqrt{-i-k^2}}{\sqrt{i-k^2}}\right)^\nu
\right]\notag\\
&+
\calO(r^{5/2-\nu}).\notag
\end{align}
We see from this form that for $1/2<\nu<1$, we allow for solutions which blow up as $r\to 0$.  They still obey prescribed boundary conditions, however; specifically, solutions for which the $r^{1/2-\nu}$ coefficient is nonzero correspond to generalized Neumann or mixed boundary conditions, as described in \cite{Ishibashi2004}.  In order to compare to their analysis, we now rewrite the coefficients in square brackets in (\ref{psialphanear0}) as
\begin{equation}\label{AdScoeffs}
\alpha_\nu \equiv
1+e^{i\alpha}\left(
\frac{\sqrt{i-k^2}}{\sqrt{-i-k^2}}
\right)^\nu,
\quad
\beta_\nu \equiv
1+e^{i\alpha}\left(
\frac{\sqrt{-i-k^2}}{\sqrt{i-k^2}}
\right)^\nu.
\end{equation}
Recalling that the chosen square roots are those with positive imaginary components, we can reduce these coefficients to
\begin{equation}
\alpha_\nu=1+\exp \left(
i\alpha-i\nu\arctan \frac{1}{k^2}
\right),
\quad
\beta_\nu=1+\exp\left(
i\alpha+i\nu\arctan\frac{1}{k^2}
\right).
\end{equation}
Both $\alpha_\nu$ and $\beta_\nu$ range from $0$ to $2$.  Consequently their ratio, $\beta_\nu/\alpha_\nu$, ranges from $0$ to $\infty$.  We can thus safely label our choice of self-adjoint extension by the this ratio, rather than the value of $\alpha$.  One reason for this choice of labelling here is that that the particular value of $\alpha$ which produces a given ratio $\beta_\nu/\alpha_\nu$ depends on $k$; this is an artifact of the Poincar\'e coordinates we have chosen to work in.

More importantly, the ratio $\beta_\nu/\alpha_\nu$ corresponds most directly to the boundary conditions implied by the choice of extension. As in \cite{Ishibashi2004}, further analysis is simplest for the case $\nu=1/2$.  We must use a different expansion for the Hankel functions valid for $\nu=1/2$, but the analysis follows exactly as in \cite{Ishibashi2004}.

In this case we find that $\beta_\nu/\alpha_\nu$ controls the ratio $\partial_r\psi/\psi$ at the boundary $r=0$.  Thus $\alpha_\nu=0$, or $\beta_\nu/\alpha_\nu=\infty$, corresponds to regular Dirichlet conditions, $\beta_\nu/\alpha_\nu=0$ corresponds to Neumann conditions, and other values correspond to mixed Robin conditions.

For more general values of $0\leq \nu <1$, \cite{Ishibashi2004} call the cases $\beta_\nu/\alpha_\nu=\infty,\, 0$ generalized Dirichlet, Neumann conditions, and all other choices are termed generalized Robin conditions.

The domain of a particular extension, as defined in (\ref{AUdomain}), does determine the boundary conditions for the time-propagated solution $\phi_t$.  That is, even though we begin with compact data $\psi_0,\, \dot{\psi_0}$, the definition of $\phi_t$ in (\ref{phitequation}) will ensure in general that future slices have instead the boundary conditions of the full domain (\ref{AUdomain}).

Recalling the definition of $\psi$ in (\ref{AdSpsidef}), we can recover the boundary conditions for $\phi_t$.  We find the behavior
\begin{equation}
\phi_t\sim \#\alpha_\nu r^{d/2-1-\nu}+ \# \beta_\nu r^{d/2-1+\nu}+\calO(r^{d/2+1-\nu}),
\end{equation}
where again $\beta_\nu/\alpha_\nu$ controls the boundary conditions, and $\#$ represent numbers not dependent on the choice of $\alpha$.  We note that not all boundary conditions may be possible for a given $k,\, \nu$; we have not tested for positivity of the extension.  In fact, \cite{Ishibashi2004} found that only a partial range of choices for $\beta_\nu/\alpha_\nu$ produce a positive extension.  We leave out these details here; the interested reader can see \cite{Ishibashi2004} for the precise ranges involved.

\section{Scalars in Lifshitz}\label{Lifsection}

We now consider the Lifshitz spacetime in $d+1$ dimensions, with metric given as:
\begin{equation}\label{LifMet}
ds^2_{d+1} = - \frac{dt^2}{r^{2z}} + \frac{d\vec{x}^2}{r^2} + \frac{dr^2}{r^2},
\end{equation}
where $x$ runs over $d-1$ dimensions. This spacetime is static and also stably causal.  The boundary in these coordinates is at $r=0$. We will not discuss the tidal singularity present at the center of the spacetime where $r=\infty$. Indeed, we will see later in equation (\ref{innerproduct2}) that the natural norm on scalar fields will require that these fields be strongly suppressed near the tidal singularity.

For $z>1$, multiplying by a conformal factor $r^{2z}$ produces a degenerate metric at $r=0$. As discussed in the introduction, this behavior is by design, since this spacetime is designed to be dual to a field theory with different scalings for time and space. For our purposes, we will privilege the time slices $\Sigma$ defined by constant $t$, which are $d$ dimensional and have a well defined non-degenerate $d-1$ dimensional boundary at $r=0$.

Before we analyze the operator $A$ as defined in (\ref{Aeqn}) for the Lifshitz spacetime, we wish to highlight the issue of the Hilbert space in question.  As mentioned above, the operator $A$ acts on the Hilbert space of functions which live on $\Sigma$ and are square integrable with respect to the measure $V^{-1} d\Sigma$.  Specifically for the Lifshitz metric (\ref{LifMet}), we find $V=r^{-z}$, and thus the appropriate measure is
\begin{equation}\label{measure}
V^{-1} d\Sigma=r^z d\Sigma=r^{z-d} \,dr \, d\vec{x}_{d-1}
\end{equation} where $d\vec{x}_{d-1}$ is the volume form for the $\mathbb{R}^{d-1}$ slice in Lifshitz geometry.
The inner product is given by
\begin{equation}\label{innerproduct}
\langle\phi_2|\phi_1\rangle=\int_\Sigma \phi_2^* \, \phi_1 r^{z-d} \, dr\, d\vec{x}_{d-1}.
\end{equation}
We can see that the set of functions included in the Hilbert space does depend on $z$.

The Klein-Gordon equation $\nabla^\mu \nabla_\mu \phi-m_0^2\phi=0$ for the metric (\ref{LifMet}) is
\begin{equation}\label{expandedKG}
-r^{2z}\partial_t^2 \phi +r^{d+z} \partial_r\left[r^{2-d-z}\partial_r\phi\right]+r^2 \vec{\nabla}\cdot \vec{\nabla} \phi-m_0^2\phi=0,
\end{equation}
where $\vec{\nabla}$ is the gradient on the $\mathbb{R}^{d-1}$ slices.
We expand $\phi$ as
\begin{align}\label{phitopsi}
\phi &= r^p \int d\vec{k} \, c_{\vec{k}}\, e^{i \vec{k}\cdot \vec{x}} \, \psi_{k} (t, r) \\
p &= \frac{d+z-2}{2}, \notag
\end{align}
where $c_{\vec{k}}$ only depends on the value of $\vec{k}$.  The $\psi_{k}$ only depend on the magnitude of the momentum $\vec{k}$, as we can infer from the form of (\ref{expandedKG}).
With this expansion we can now rewrite (\ref{expandedKG}) in the form prescribed in (\ref{Aeqn}):
\begin{equation}\label{AKG}
\partial_t^2\psi_k (t, r) =-A \,\psi_k (t, r),
\end{equation}
where the operator $A$ is given by
\begin{equation}\label{Adef}
A=-r^{2(1-z)}\left[\partial_r^2-\frac{1}{r^2}\left(m_0^2+p(p+1)\right) -k^2\right].
\end{equation}
We now would like to write the norm (\ref{innerproduct}) in terms of the $\psi_k$.  The $e^{i\vec{k}\cdot\vec{x}}$ are orthogonal but not normalizable.  Since we are interested in the behavior at the boundary $r=0$, we will only study the behavior of the $\psi_k (t,r)$ pieces, assuming that we can arrange wave packets which are normalizable in the $\vec{x}$ directions.
In terms of $\psi_k (t,r)$, the relevant portion of the norm (\ref{innerproduct}) becomes
\begin{equation} \label{innerproduct2}
\langle\phi_2|\phi_1\rangle_r = \int dr\, r^{2z-2} \, \psi_{2,k}^*(t,r) \, \psi_{1,k}(t,r).
\end{equation}

We now have set up our problem: we want to study boundary conditions at $r=0$ for $\psi_k (t,r)$ which solve (\ref{AKG}) with $A$ as in (\ref{Adef}) and are finite under the norm (\ref{innerproduct2}).  In the remainder of this section, we will study the behavior of solutions to the complex eigenvalue equation (\ref{Aieqn}), then define the self adjoint extensions $A_U$ of this $A$, and finally study which of these extensions are finite and what boundary conditions they correspond to.

\subsection{Finding the solution spaces $\mathcal{K}_\pm$}

We now return to the equation (\ref{Aieqn}) with operator $A$ as in (\ref{Adef}) in order to find the solution spaces $\mathcal{K}_\pm$.  Since we only want to understand which solutions to (\ref{Aieqn}) are normalizable under (\ref{innerproduct2}), we actually do not need to solve for the full eigenfunctions $\psi_\pm$; instead we will only need to understand their behavior near the boundary. For ease of notation, we will actually analyze the behavior of the eigenfunctions for which $A\psi=\lambda \psi$  for generic rational values of $z$, and general complex eigenvalues $\lambda$. We also drop the subscript $k$.
We rewrite this eigenequation as
\begin{equation}\label{Rniceform}
r^2 \partial_r^2 \psi-\left(\nu^2-\frac{1}{4}\right)\psi-k^2 r^2 \psi+\lambda r^{2z}\psi =0,
\end{equation}
where $\nu$ is given by
\begin{equation}\label{nudef}
\nu^2\equiv m_0^2+\left(\frac{d+z-1}{2}\right)^2.
\end{equation}
Note that $\nu^2=0$ corresponds to the equivalent of the Breitenlohner-Freedman bound in this space, $m_0^2=-((d+z-1)/2)^2$.  Considering only $\nu^2\geq 0$ ensures the operator $A$ is positive on our Hilbert space.%
\footnote{Usually fields below the Breitenlohner-Freedman bound are associated with instabilities; \cite{Moroz2009} indicates that the situation is different for Schroedinger spacetimes.  Lifshitz apparently is more similar to AdS; we think the positivity bound on $A$ for $\nu^2$ here indicates that massive scalars below the BF bound in Lifshitz will suffer from an instability.}
Consequently, we will only consider $\nu \geq 0$ in our subsequent analysis.

Equation (\ref{Rniceform}) is a second order linear differential equation with two singular points, $r=0$ and $r=\infty$.  As such, there are two linearly independent solutions for any given $\lambda$.  We are only interested in solutions which are square integrable under the measure (\ref{innerproduct2}); thus we will examine the behavior of solutions near each of the singular points for such square integrability.

The singular point at $r=0$ is regular as long as $z\geq 0$, which includes the range we study: $z>1$.  Considering a power series solution $\psi=\sum_n a_n r^{b+n}$ to (\ref{Rniceform}), the equation becomes
\begin{align}
0=& a_0 \left[b(b-1)-\nu^2+\frac{1}{4}\right]r^b+a_1\left[(b+1)b-\nu^2+\frac{1}{4}\right]r^{b+1}\notag\\
&+ \sum^\infty_{n=2}\left[a_n(b+n)(b+n-1)-(\nu^2-\frac{1}{4})a_n-\vec{k}^2a_{n-2}\right]r^{b+n}\\
\notag &+  \sum^\infty_{n=2z}\left[\lambda a_{n-2z}r^{b+n}\right].
\end{align}
for half-integer $z$.
The coefficient for each power of $r$ must vanish separately, so the indicial equation becomes
\begin{equation}\label{indicial}
\left[b(b-1)-\nu^2+\frac{1}{4}\right]=0.
\end{equation}
For rational $z$ which are not half integers, we may derive the same effective indicial equation by changing variables to $\rho=r^{1/m}$ for $m$ such that $mz$ is an integer.  This equation gives the leading behaviors of the two solutions for $\psi$ near $r=0$ as
\begin{equation}\label{Rpmdef}
\psi_{\uparrow} \propto r^{b_+}, \qquad \psi_{\downarrow} \propto r^{b_-}, \qquad b_{\pm}=\frac{1}{2}\pm \nu,
\end{equation}
assuming $\nu$ is not itself zero or a half integer. $\nu$ must be nonnegative in order to assure positivity of the operator $A$ on the desired Hilbert space.

When $\nu$ is zero, the two roots of the indicial equation (\ref{indicial}) coincide, and thus the leading behaviors become $r^{1/2}$ and $r^{1/2}\log r$. For $\nu$ some nonzero half integer, the two roots will differ by the integer $2\nu$.  Here, the leading behavior as $r$ approaches zero will be $r^{1/2-\nu}$, but there may be logarithmic terms starting with $r^{1/2+\nu}\log r$ in the full expansion.

A solution $\psi(r)$ will be square integrable near zero  with measure as in (\ref{innerproduct2}) when the leading behavior of $\psi^2 r^{2z-2}$ is $r^a$ for $a>-1$.  The solution $\psi_\uparrow$ has behavior near $r=0$ such that
\begin{equation}\label{Rupintegrable}
r^a=r^{2z-2+2\nu+1}; \qquad\text{square integrable if }\nu>-z.
\end{equation}
As we are working only with $\nu\geq 0$, this solution will always be square integrable near zero.

For the solution $\psi_\downarrow$ when $\nu$ is strictly greater than zero, we find
\begin{equation}\label{R-integrable}
r^a=r^{2z-2-2\nu+1}; \qquad\text{square integrable if } -\nu>-z,
\end{equation}
as the possible logarithmic behavior is subleading. For the case $\nu=0$, we wish to study the leading logarithmic behavior
\begin{equation}
\label{R-fornuzero}
r^{2z-2}r^{2(1/2)}\log{r}^2,
\end{equation}
which is integrable near zero for any $z>1$.

These results indicate that for $0\leq\nu<z$, both solutions $\psi_{\uparrow,\downarrow}$ are square integrable near zero. For $\nu\geq z$, only the $\psi_\uparrow$ solution remains in the Hilbert space, regardless of the eigenvalue $\lambda$ under consideration.

In order to find the possible square integrable solutions to the eigenvalue equation (\ref{Rniceform}), we must also consider the behavior of these solutions near the irregular singular point at infinity.  Again, as near $r=0$, we expect two solutions each with different behavior.

The point at $r=\infty$ is an irregular singular point with rank $z$.  Thus there must exist formal solutions, or asymptotic series, of the form
\begin{equation}\label{asympseriesR}
\psi=\exp\left[\sum_{n=1}^{z} c_n r^n\right]Q(r),\qquad Q(r)=\sum_{l=0}^\infty a_l r^{b-l}.
\end{equation}
For more details on such solutions, see e.g. \cite{Goldstein}.  Plugging this form into (\ref{Rniceform}) and dividing by $r^2 \exp\left[\sum_{n=1}^{z} c_n r^n\right]$ gives
\begin{align}\label{asympseriesReqn}
0= &\partial_r^2 Q+
2\sum_{n=1}^z n c_n r^{n-1}\partial_r Q\\
& +\left[\sum_{n=1}^z n (n-1)c_n r^{n-1}+\left(\sum_{n=1}^z n c_n r^{n-1}\right)^2-\left(\nu^2-\frac{1}{4}\right)r^{-2}-\vec{k}^2+\lambda r^{2z-2}\right]Q.\notag
\end{align}
We will start by setting to zero the coefficient of the highest power of $r$, as these are the most dominant terms near $r=\infty$.  The largest power in $Q$ is $r^b$; for $\partial_r Q$ the largest power is $r^{b-1}$, and for $\partial_r^2 Q$ it is $r^{b-2}$.  Thus the largest overall power present in (\ref{asympseriesReqn}) is $r^{2z-2+b}$, arising from the $\lambda$ term and the highest term in the squared sum in the second line.  Specifically this contribution is
\begin{equation}\label{largestrpower}
r^{2z-2+b}\left(z^2 c_z^2 +\lambda\right)=0,
\end{equation}
and it must vanish independently. Note that if $z=1$, the $-\vec{k}^2$ term is of the same order and would contribute; this is precisely the AdS case studied above.  Here we will restrict our attention to $z>1$.
As for the behavior near $r=0$ for rational $z$, we can find the behavior for rational, noninteger $z$ here by changing variables to $\rho=r^{1/m}$ for $m$ such that $mz$ is an integer. We again recover equation (\ref{largestrpower}) for the coefficient of the leading power in the exponent of $\psi$.

The behavior of $\psi$ near infinity is controlled by $c_z$, which must be a root of equation (\ref{largestrpower}). Labelling the two roots of this equation $c_{z,\pm}$, we can write the two allowed behaviors of $\psi$ near infinity as
\begin{align}\label{Rnearinf}
\psi_1 &\propto \exp\left[c_{z,+}r^z+\calO(r^{z-1})\right]\left(r^b+\calO(r^{b-1})\right)\\
\psi_2& \propto \exp\left[c_{z,-}r^z+\calO(r^{z-1})\right]\left(r^b+\calO(r^{b-1})\right).
\notag
\end{align}
If the real part of $c_{z,\pm}$ is positive, the solution blows up exponentially and will not be square integrable under the measure (\ref{measure}).  Conversely, if the real part of $c_{z,\pm}$ is negative, then the solution is exponentially damped and will be square integrable.

Since we are interested in the solution space for (\ref{Aieqn}), we now examine the roots of (\ref{largestrpower}) for $\lambda=\pm i$.  We find
\begin{align}\label{czforlambdaimag}
\lambda=i & \rightarrow c^{i}_{z,\pm}=\pm \left(\frac{1+i}{z\sqrt{2}}\right),\\
\lambda=-i & \rightarrow c^{-i}_{z,\pm}=\pm \left(\frac{1-i}{z\sqrt{2}}\right).
\notag
\end{align}
In either case, the solution $\psi_2$ built on $c_{z,-}$ is square integrable, and $\psi_1$ built on $c_{z,+}$ is not, for every rational $z>1$.

We are searching for solutions to (\ref{Aieqn}) which are square integrable under measure (\ref{measure}) over their full range.  The only candidate solution near $r=\infty$ is $\psi_2$.  This solution will be a linear combination of those near $r=0$, denoted $\psi_{\uparrow,\downarrow}$ as in (\ref{Rpmdef}):
\begin{align}\label{linearcombo}
\psi_2=C_\uparrow \psi_\uparrow+ C_\downarrow \psi_\downarrow,
\end{align}
for constants $C_\uparrow$, $C_\downarrow$. For $0\leq \nu <z$, both $\psi_\uparrow$ and $\psi_\downarrow$ are appropriately square integrable near $r=0$, so in this range the solution $\psi_2$ itself will be square integrable with measure (\ref{measure}) from $0\leq r< \infty$.

Conversely, for $\nu\geq z$, the solution $\psi_\downarrow$ grows too fast near $r=0$.  Thus $\psi_2$ will not be a valid square integrable solution to (\ref{Aieqn}) unless $C_\downarrow=0$, which will not occur generically.

In summary, for $0\leq \nu <z$, one generating solution to (\ref{Aieqn}) for each of $\pm i$ is square integrable.  Since we can multiply this solution by any phase, we have a one-dimensional set of solutions for each of $\pm i$, and thus we will find a one-dimensional alternative set of boundary conditions available for modes with this range of $\nu$.  For $\nu\geq z$, no such solution is available, and only one type of boundary condition is sensible.

\subsection{Boundary conditions for $\psi$ when $0<\nu<z$}

We have shown that there is a one-dimensional space of possible boundary conditions for scalar operators whose mass satisfies $\nu<z$.  We now wish to explore what boundary conditions these are. The added complication of the factor $z$ in the equation (\ref{Rniceform}) will prevent us from finding as explicit a form for the boundary conditions as we could in the AdS case.  In particular, we will not be able to find the coefficients $C_\uparrow,\, C_\downarrow$ which determine the precise behavior of the function $\psi_2$ near the boundary $r=0$.  However, although we will not be able to find these values, we can still discuss the available behaviors near the boundary.

As in the AdS case, the maps $U$ from the spaces $\cal{K}_\pm$ for square integrable solutions $\lambda=\pm i$ are of the form (\ref{phasemap}) and thus just introduce a phase. Consequently we again define $\psi_\alpha=\psi_{2,\lambda=i}+e^{i\alpha} \psi_{2,\lambda=-i}$.

In (\ref{Rpmdef}), we find the behaviors of $\psi_{\uparrow,\downarrow}$ near $r=0$ to be $1/2 \pm \nu$. For $\nu$ not a half integer, we can use these behaviors to write the behavior of $\psi_\alpha$ as
\begin{equation}\label{psialphaznear0}
\psi_\alpha \sim \sum_{n=0}^\infty (b_{n,\alpha} r^{1/2-\nu+n} + c_{n,\alpha} r^{1/2+\nu+n}),
\end{equation}
where the coefficients $c_{n,\alpha},\, b_{n,\alpha}$ depend on the choice of map, and control the choice of boundary conditions.  Unfortunately our inability to match the solutions $\psi_{\uparrow,\downarrow}$ near the boundary $r=0$ to the solutions $\psi_{1,2}$ near the tidal singularity $r=\infty$ means we cannot find the ranges of the coefficients $c_{n,\alpha},\, b_{n,\alpha}$.  Similarly we cannot test within this range to find which boundary conditions correspond to positive choices of extension.

However, we still expect that the choice of extension, or of $\alpha$, corresponds to a choice for the ratio $c_{0,\alpha}/b_{0,\alpha}$.  We still define $\phi_t$ via (\ref{phitequation}) and expect its boundary conditions to explore the full domain of the choice of extension determined by $c_{0,\alpha}/b_{0,\alpha}$.

Using the relationship between $\phi$ and $\psi$ as in (\ref{phitopsi}), we find the behavior of $\phi_t$ near $r=0$ to be
\begin{equation}
\phi_t \sim \sum_{n=0}^\infty (b_{n,\alpha} r^{(d+z-1)/2-\nu+n}+c_{n,\alpha} r^{(d+z-1)/2+\nu+n}).
\end{equation}
The leading term here goes like $r^{-\nu+(d+z-1)/2}$.  As choices of boundary conditions are available for scalar fields whose $\nu$ satisfies $0\leq \nu <z$, we find that the exponent of this term will be bounded below by $(d-z-1)/2$.   Thus, for Lifshitz spacetimes in $d+1$ dimensions where $z+2 > d+1$, one can choose that some scalar field will have a blow-up near the boundary $r=0$.

This behavior could occur when the mass is as far above the Breitenlohner-Freedman bound as we can achieve and still have a choice of boundary condition behavior. We found that a choice of boundary conditions is possible whenever $0\leq\nu <z$, or when $m_0^2$ satisfies
\begin{equation}\label{massrange}
-\left(\frac{d+z-1}{2}\right)^2 \leq m_0^2 < z^2 -\left(\frac{d+z-1}{2}\right)^2.
\end{equation}
We expect that scalar fields with values of $m_0^2$ near the top end of this range will have a boundary condition choice resulting in growth near $r=0$, as long as the Lifshitz spacetime under consideration has $z+2>d+1$.  In fact, spacetimes above this limit have been considered previously but been found to have other sicknesses; see \cite{Ross2011} and references for details.

The second series in $\phi_t$ controlled by $c_{0,\alpha}$ begins down by a power $r^{2\nu}$.  Again, as we can pick alternate boundary conditions for values of $\nu$ up to $z$, we can see that these conditions correspond to a further generalization of the Neumann, Dirichlet and Robin conditions.  In particular in the case $\nu=n+1/2$, we might guess that the alternate boundary conditions here relate $\psi$ to the derivative $\partial^n_r \psi$.

\section{Summary and Future Directions}\label{conclusions}

We have studied the boundary conditions available for Lifshitz spacetimes which are solutions to Einstein gravity plus unspecified matter.  We find a one parameter family of boundary condition types for scalar fields with effective masses satisfying (\ref{massrange}).  In the AdS case, we label these conditions by the ratio $\alpha_\nu/\beta_\nu$, where $\alpha_\nu$ multiplies the leading behavior, and $\beta_\nu$ the subleading behavior.  For the Lifshitz case, we define similarly $b_{0,\alpha},\, c_{0,\alpha}$. A choice of ratio $\alpha_\nu/\beta_\nu$, or $b_{0,\alpha}/c_{0,\alpha}$, corresponds to a choice of self-adjoint extension or boundary condition type. These types range from a generalized Dirichlet condition, through a family of mixed conditions, to a generalized Neumann condition. We summarize our results in Table \ref{summarytable}.
\begin{table}[h]
\centering
\begin{tabular}{|c|c|c|c|}
\hline

& $\nu=0$
& $0 <\nu < z$
& $\nu>z$
\\
\hline
mass
&$ m_0^2 = -\left(\frac{d+z-1}{2}\right)^2$
&$-\left(\frac{d+z-1}{2}\right)^2\leq m_0^2 < z^2 -\left(\frac{d+z-1}{2}\right)^2$
&$ m_0^2 \geq z^2-\left(\frac{d+z-1}{2}\right)^2$
\\
\hline
$\phi_t$ leading order
& $r^{(d+z-1)/2} \log r$
& $r^{-\nu+(d+z-1)/2}$
&
\\
\hline
$\phi_t$ subleading
& $r^{(d+z-1)/2}$
& $r^{\nu+(d+z-1)/2}$
& $r^{\nu+(d+z-1)/2}$
\\
\hline
\end{tabular}
\caption{Summary of results: the leading order behavior of $\phi_t$ is controlled by $\alpha_\nu$ for AdS, or $b_{0,\alpha}$ for the Lifshitz case. The subleading is controlled by $\beta_\nu$ or $c_{0,\alpha}$. Thus the boundary conditions are set by the ratio $\alpha_\nu/\beta_\nu$, or $b_{0,\alpha}/c_{0,\alpha}$.  For $\nu>z$, no choice is available.}
\label{summarytable}
\end{table}

Additionally, following \cite{Ishibashi2004}, we can define a conserved, positive energy for each choice of boundary condition, parameterized here by $-\pi<\alpha<\pi$:
\begin{equation}
E(\psi) \equiv \int dr \, r^{2z-2}\dot{\psi}^* \, \dot{\psi}+
\int dr \, r^{2z-2} \psi^* \left(
A\psi+ i\psi_{2,\lambda=i}-ie^{i\alpha}\psi_{2,\lambda=-i}
\right),
\end{equation}
where $A$ is given in (\ref{Adef}), and we have used the definitions in (\ref{Aextension}) with $\psi_{2,\lambda}$ as in (\ref{Rnearinf}).

Below the mass range (\ref{massrange}) the Laplacian operator is not positive; the lower end corresponds to a Breitenlohner-Freedman bound.  Above this mass range, we cannot choose the type of boundary condition; only a Dirichlet-type condition is available.  Importantly, we are able to reproduce the results for scalar fields found in \cite{Ishibashi2004} for the $z=1$, that is the AdS, case, albeit in Poincar\'e coordinates.

The authors of \cite{Ishibashi2004} also found alternate boundary condition options for gravity and gauge fields.  The authors accomplished this by expanding these fields in modes in global coordinates, resulting in a set of effective scalar equations with an extra parameter $\sigma$.  As Lifshitz spacetime does not have a good set of global coordinates, we have not done a similar expansion.

We have also not investigated the effect of the tidal singularity (at $r=\infty$ in our coordinates). The method of exploring boundary conditions proposed in \cite{Ishibashi2003} should work for curvature singularities as well as for boundaries at infinity; it would be interesting to study the method for tidal singularities as well.  Although we have not done this analysis, we already see that the solutions to (\ref{Aieqn}) which are square integrable are in fact exponentially damped as $\exp (-r^z)$ in the region $r\to \infty$.

It would also be interesting to compare the alternate boundary conditions we have found here to previous work on holography in Lifshitz spacetimes.  Particularly, we wonder about the relation to a certain mixed boundary condition required in \cite{Ross2011}.  Additionally we have only considered Lifshitz spacetimes as solutions of Einstein gravity; interesting recent work \cite{Griffin2012} suggests that Ho\v{r}ava-Lifshitz gravity could be a natural ground for holographic duals to Lifshitz field theories.

One can also ask what these alternate boundary conditions correspond to in the
putative dual Lifshitz field theory.  In the AdS case, a choice of mixed boundary condition corresponds to adding a multitrace operator in the field theory, as studied in ``designer gravity'' as in \cite{Amsel2006}.  It would be interesting to develop a similar ``designer Lifshitz gravity'' story.

\vspace{.25in}

\hspace{-.25in}\textbf{Acknowledgements}

I would like to thank Ibrahima Bah for significant collaboration in earlier stages of this project. I also thank Balt van Rees, David Garfinkle, Dionysius Anninos, Don Marolf, Jim Liu, Monica Guica, Simon Ross, and Tom Hartman for useful and inspiring conversations. CK is supported in part by DoE Grant DE-SC0007859.

\bibliographystyle{utcapsnourl}
\bibliography{NonConformalBoundaries.bib}

\end{document}